\documentclass[twocolumn]{article}

\usepackage{stfloats}
\usepackage{ctable}
\usepackage{amsmath}
\usepackage{amsfonts}
\usepackage{amssymb}
\usepackage{graphicx}%
\usepackage{algorithm}
\usepackage{algorithmic}
\usepackage{amsmath}
\usepackage{dsfont}
\usepackage{pxfonts}
\usepackage{alltt}
\usepackage{tabularx}

 \usepackage{graphicx}
 \usepackage{amsmath}
 \usepackage{amssymb}
\usepackage{times}
 \usepackage{algorithm}
\usepackage{algorithmic}

\usepackage[standard]{ntheorem}

\usepackage{amsfonts}
\usepackage{graphicx}%
\usepackage{algorithm}
\usepackage{algorithmic}
\usepackage{amsmath}

 \usepackage{amsmath}
 \usepackage{amssymb}
\usepackage{times}
 \usepackage{algorithm}
\usepackage{algorithmic}

\begin{document}

\title{Two Security Layers for Hierarchical Data Aggregation in Sensor Networks}

\author{Jacques M. Bahi, Christophe Guyeux, and Abdallah Makhoul}
\maketitle

\begin{abstract}

Due to resource restricted sensor nodes, it is important to minimize the amount of data transmission among sensor networks. To reduce the amount of sending data, an aggregation
approach can be applied along the path from sensors to the
sink. However, as sensor networks are often deployed in untrusted and even hostile environments, sensor nodes are prone to node compromise attacks. Hence, an end-to-end secure aggregation approach is required
to ensure a healthy data reception. In this paper, we propose two layers for secure data aggregation in sensor networks. Firstly, we provide an end-to-end encryption scheme that supports operations over cypher-text. It
is based on elliptic curve cryptography that exploits a smaller key
size, allows the use of higher number of operations on
cypher-texts, and prevents the distinction between two identical texts
from their cryptograms. Secondly, we propose a new watermarking-based authentication that enables sensor nodes to ensure the identity of other nodes they are communicating with. Our experiments show that our hybrid approach of secure data aggregation enhances the security, significantly reduces computation and communication overhead,
and can be practically implemented in on-the-shelf sensor platforms.
\end{abstract}

\section{Introduction}

A typical sensor network is expected to consist of a large number of sensor nodes
deployed randomly in a large scale. Usually, these nodes have limited
power, storage, communication, and processing capabilities, making
energy consumption an issue.

A major functionality of a sensor node is to measure environmental
values using embedded sensors, and transmit it to a base station
called "sink". The sensed data needs to be analyzed, which
eventually serves to initiate some action. Almost this analysis
presumes computation of the maximum, minimum, average, \emph{etc}. It
can be either done at the base station or by the nodes themselves, in
a hierarchical scenario. In order to reduce the amount of data to be
transmitted to the sink, it is beneficial that this analysis can be
done over the network itself. To save the overall energy resources of
the network, it is agreed that the sensed data needs to be aggregated
on the way to its final destination. Sensor nodes send their values to
certain special nodes, i.e., aggregators. Each aggregator then
condenses the data prior to sending it on. In terms of bandwidth and
energy consumption, aggregation is beneficial as long as the
aggregation process is not too central processing unit (CPU) intensive. The aggregators can
either be special (more powerful) nodes or regular sensors nodes.

At the same time, sensor networks are often deployed in public or
otherwise untrusted and even hostile environments, which prompts a
number of security issues (e.g., key management, privacy, access
control, authentication, \emph{etc.}). Then, if security is a
necessary in other (e.g., wired or MANET) types of networks, it is
much more so in sensor networks. Actually, it is one of the most
popular research topic in this field and many advances have been reported on in
recent years.

From the above observations, we can notice the importance of a
cooperative secure data aggregation in sensor networks. In other
terms, after the data gathering and during transmissions to the base
station, each node along the routing path cooperatively integrates and
secures the fragments messages. Therefore, secure data aggregation protocols require sensor
nodes to encrypt or authenticate any sensed data
prior to its transmission, implement data
aggregation at every intermediate node (without decryption), and prefer data to be decrypted
by the sink so that energy efficiency is maximized.

The benefit and vulnerability, as well as the need to secure
in-network aggregation, have been identified by numerous schemes in
the literature. One approach~\cite{76} proposed a secure information
aggregation protocol to answer queries over the data acquired by the
sensors. Even though their method provided data authentication to
guarantee secrecy, the data still sent in plain text format, which removes
the privacy during transmission. Another one~\cite{77} proposed a
secure energy efficient data aggregation (ESPDA) to prevent redundant
data transmission in data aggregation. Unlike conventional techniques,
their scheme prevents the redundant transmission from sensor nodes to
the aggregator. Before transmitting sensed data, each sensor transmits
a secure pattern to the aggregator. Only sensors with different data
are allowed to transmit their data to the cluster-head. However, since
each sensor at least needs to transmit a packet containing a pattern
once, power cannot be significantly saved. In addition, each sensor
node uses a fixed encryption key to encrypt data, which can lead to severe security flaws. In~\cite{78}, the authors presented a
secure encrypted-data aggregation scheme for wireless sensor
networks. The idea is based on eliminating redundant sensor readings
without using encryption and maintains data secrecy and privacy during
transmission. This scheme saves energy on sensor nodes but still do not
guarantee the privacy of sent data.

In this paper, we provide for the first time an hybrid approach for secure data aggregation in sensor networks. Firstly, our approach ensures that secrecy of sensed data is never disclosed to unauthorized parties by providing a secure homomorphic cypher-system that allows efficient aggregation of encrypted data. We show that our encryption method allows many operations over cypher-texts that prevents data decryption at intermediate nodes (aggregators) and reduces energy consumption. Secondly, we extend our homomorphic secure data aggregation level to two layers hierarchical data aggregation protocol, by including a watermarking-based authentication level. To assess the practicality of
our technique, we evaluate it and compare it to existing
cypher-system. The obtained results show that we significantly reduce
computation and communication overhead as well as our secure aggregation
method can be practically implemented in on-the-shelf sensor
platforms.

The rest of this paper is organized as follows.
After having recalled some previous related work in the fields of data confidentiality and authentication, the next section introduces our two security layers.
In Section \ref{section:secure data} the first one, namely the secure data aggregation using an almost fully homomorphic cryptosystem over elliptic curves, is presented in detail.
Its security is evaluated qualitatively and through experiments in Section \ref{section:evaluation}.
In the next section, our second complementary approach for security in WSN is proposed.
This is a nodes authentication protocol based on information hiding security field.
Advanced notions of security are taken from this field and translated in WSN terms.
Then an existing authentication scheme is evaluated and improved in Section \ref{section:Zhang}.
Finally, in Section \ref{section:dhci}, a new secure authentication method based on watermarking is proposed and evaluated. 
This research work ends by a conclusion section, where our contribution is summarized and intended future work is presented.

\section{Security in sensor networks}
\label{section:Security}

Because sensor networks may interact with sensitive
data and be deployed in hostile unattended environments, it is imperative
to protect sensitive information transmitted by sensor nodes. Moreover, wireless sensor networks introduce severe resource constraints due
to their lack of data storage and power. Therefore, they have security
problems that traditional networks (computer security) do not face and there are many security considerations that
should be investigated. In this paper, we treat the
essential security requirements that are raised in a wireless
sensor network environment, mainly: data confidentiality, node authentication, and how they relate with data aggregation process.

\subsection{Data confidentiality}

In critical applications, data confidentiality ensures that secrecy of transmitted data is never disclosed to unauthorized parties. Therefore, it is very important to build secure channels between sensor networks. The standard technique for keeping sensitive data secret is to encrypt them such that only intended receivers can realize decryption, hence achieving confidentiality.

Data encryption becomes necessary in sensor networks when this type of
sensors can be subject of many types of attacks~\cite{68}. Without
encryption, adversaries can monitor and inject false data into the
network. In a general manner the encryption process is done as
follows: sensor nodes must encrypt data on a hop-by-hop basis. An
intermediate node (\emph{i.e.}, aggregator) possessing the keys of all
sending nodes, decrypts the received encrypted value, aggregates all
received values, and encrypts the result for transmission to the base
station. Though viable, this approach is fairly expensive and
complicated, due to the fact of decrypting each received value before
aggregation, which generates an overhead imposed by key
management and prevents end-to-end data confidentiality.

Some privacy homomorphism based researches 
have been proposed recently~\cite{69, 70, 71} that, without
participating in checking, the aggregators can directly aggregate the
encrypted data. The problem of aggregating encrypted data in sensor networks was
introduced in~\cite{70} and further refined in~\cite{69}. The authors
propose to use homomorphic encryption schemes to enable arithmetic
operations over cypher-texts that need to be transmitted in a multi-hop
manner. However, these approaches provide a higher level of system
security, since nodes would not be equipped with private keys, which
would limit the advantage gained by an attacker compromising some of
the nodes. Unfortunately, existing privacy homomorphisms used for data
aggregation in sensor networks have exponential bound in
computation. For instance, Rivest
Shamir Adleman (RSA) based cryptosystems~\cite{72,73} are used, which
require high CPU and memory capabilities to perform exponential
operations. It is too computationally expensive to implement in
sensor nodes. Moreover, the expansion in bit size during the
transformation of plain text to cypher-text introduces costly
communication overhead, which directly translates to a faster
depletion of the sensors energy. On the other hand and from security
viewpoint, the cryptosystems~\cite{83} used in these approaches were
cryptanalized~\cite{81,82}, which means they cannot guarantee anymore high
security levels.

In this paper we try to relax the statements above by investigating
elliptic curve cryptography that allows feasible and suitable data
aggregation in sensor networks beside the security of homomorphisms
schemes. First of all, our proposed scheme for secure data aggregation in sensor
networks is based on a cryptosystem that has been proven safe and has
not been cryptanalyzed. Indeed, it is known to be the sole secure and almost fully
homomorphic cryptosystem usable now.
Another property that enforces the security
level of such an approach is coming from the fact that, as it is the case
in ElGamal cryptosystem, for two identical messages it generates two
different cryptograms. This property suggested fundamental for security
in sensor networks \cite{80,78,999}, to the best of our knowledge, was not
addressed in previous homomorphism-based security data aggregation
researches. Beside all these properties and due to the use of elliptic
curves, our approach saves energy by allowing nodes to encrypt and
aggregate data without the need of high computations. Lastly, the scheme
we use allows more aggregations types over cypher data than the
homomorphic cryptosystems used until now.
This approach is detailed in Section
\ref{section:secure data} and is evaluated in Section
\ref{section:evaluation}.

\subsection{Node authentication}

In wireless sensors networks an adversary can change
the whole packet stream by injecting additional packets. Therefore, the receiver
needs to ensure that the data used in any decision-making process originates
from the correct source. Without
authentication, an adversary could masquerade a node,
thus gaining unauthorized access to resource and sensitive
information, and interfering with the operation of other
nodes. Moreover, a compromised node may send data to
its data aggregator under several fake identities so that
the integrity of the aggregated data is corrupted. Node authentication enables a sensor node to ensure the identity
of the peer node it is communicating with. In the case
of only two-nodes communication, authentication can be achieved through a
purely symmetric key cryptography: the sender and the receiver share a secret key
to compute the message authentication code (MAC) of all communicated
data.

In-network processing presents a critical challenge for data authentication in wireless sensor networks. Current schemes relying on MAC cannot provide natural support for this operation, because a MAC computation is a very energy-consuming operation. Additionally, even a slight modification to the data invalidates the MAC.

The authors in~\cite{86} propose a key-chain distribution system for their
$\mu$TESLA secure broadcast protocol. The basic idea of the $\mu$TESLA
system is to achieve asymmetric cryptography by delaying the disclosure of
the symmetric keys. In this case a sender will broadcast a message generated
with a secret key. After a certain period of time, the sender will disclose the
secret key. The receiver is responsible for buffering the packet until the secret
key has been disclosed. After disclosure the receiver can authenticate the
packet, provided that the packet was received before the key was disclosed. One major limitation of $\mu$TESLA is that some initial information must be unicast
to each sensor node before authentication of broadcast messages can begin.

In~\cite{Zhang2008658} a new way to achieve authentication through wireless sensor networks is introduced.
It is based on digital watermarking and proposes an end-to-end, statistical approach for data authentication that provides inherent support for in-network processing.
In this scheme, authentication information is modulated as a watermark and superposed on the sensory data at the sensor nodes.
The key idea formerly presented in~\cite{Zhang2008658} is to visualize the sensory data at a certain time snapshot as an image.
Each sensor node is viewed as a pixel and its value corresponds to the gray level of the pixel.
Due to this equivalence, information hiding techniques can be used to authenticate a wireless sensor network.

In some well-defined situations, the watermarked data can be aggregated by the intermediate nodes without incurring any in route checking.
In this context, aggregation is for instance related to DCT or DWT compression, that is, to any operation over images that is able to reduce their weights without removing the watermarks.
Upon reception of the sensory data, the sink is able to authenticate the data by finding and validating the watermark, thereby detecting whether the data has been illegitimately altered.
In this way, the aggregation-survivable authentication information is only added at the sources and checked by the data sink, without any involvement of intermediate nodes.
To realize such an authentication, the authors of \cite{Zhang2008658} propose to use a data hiding scheme based on spread spectrum techniques.
In their proposal, "each sensor node embeds part of the whole watermark into its sensory data, while leaving the heavy computational load of watermark detection at the sink".
Moreover, as stated before, their scheme supports in-network aggregation.
Such an approach, its issues and security consequences, and how to improve their scheme, are detailed in the last sections of this research work.


\section{Tree-based data aggregation}

\begin{figure}[h]
\centering
        \includegraphics[height=6 cm]{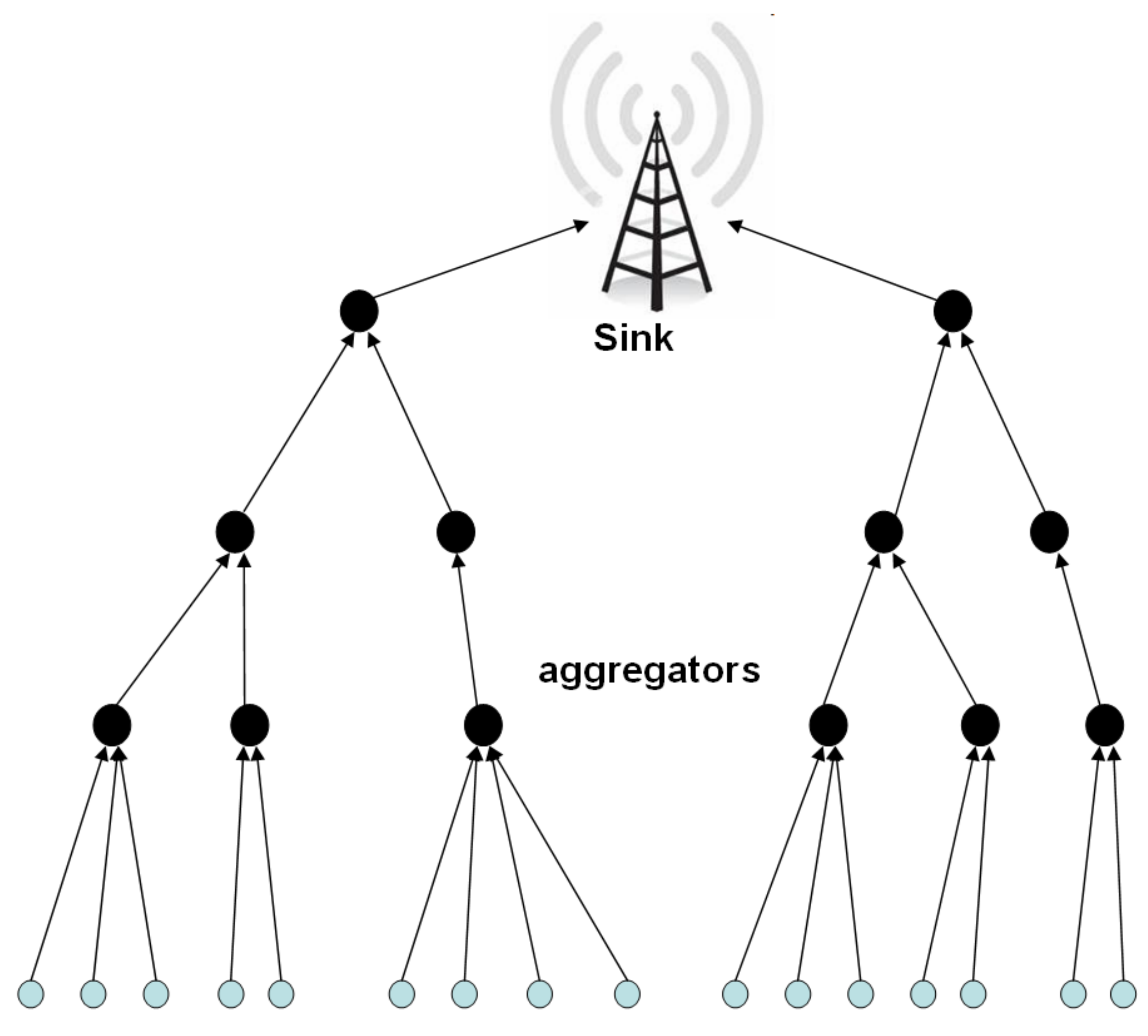}
\caption{Tree-based data aggregation in sensor networks}
\label{fig:treeagreg1}
\end{figure}

Data aggregation schemes aim to combine and summarize data packets of
several sensor nodes so that amount of data transmission is
reduced. An example of data aggregation schemes is the tree based data aggregation protocol as presented in
Figure~\ref{fig:treeagreg1} where sensor nodes collect information from a
region of interest. When the user (sink) queries the network, instead
of sending each sensor node's data to the base station, aggregators
collect the information from its neighboring nodes, aggregates them, and send the aggregated data to the base
station over a multihop path.

The main objective of data aggregation is to increase
the network lifetime by reducing the resource
consumption of sensor nodes, especially the battery energy and
bandwidth. While increasing network lifetime, data
aggregation protocols might take into account an important quality of service metric: the security. Therefore, encryption of the sensed data before its
transmission becomes necessary and it is preferable to decrypt the
data only at the base station level. In the next section, we present our model for sensor data encryption compliant with this requirement.

\section{Sensor data encryption using fully homomorphic cryptosystem}
\label{section:secure data}
In this work, we are primarily concerned with data privacy in sensor
networks. Our goal is to prevent attackers from gaining any
information about sensor data. Therefore, ensuring an end-to-end
privacy between sensor nodes and the sink becomes problematic. This is
largely because popular and existing cyphers are not additively
homomorphic. In other words, the summation of encrypted data does not
allow for the retrieval of the sum of the plain text values. Moreover,
privacy existing homomorphisms have usually exponential bound in
computation. To overcome this problem, in our model we propose a
security scheme for sensor networks using elliptic curve based
cryptosystem. We show that our model permits many operations on
encrypted data and does not demand high sensor capabilities and
computation.

\subsection{Operations over Elliptic Curves}
\label{sec:ECC}
In this section, we give a brief introduction to elliptic curve
cryptography. The reader is referred to~\cite{74} for more details.


Elliptic curve cryptography (ECC) is an approach to public-key
cryptography based on the algebraic structure of elliptic curves over
finite fields~\cite{74}. Elliptic curves used in cryptography are
typically defined over two types of finite fields: prime fields
$\mathbb{F}_p$, where $p$ is a large prime number, and binary
extension fields $\mathbb{F}_{2^m}$~\cite{79}. In our paper, we focus
on elliptic curves over $\mathbb{F}_p$. Let $p > 3$, then an elliptic
curve over $\mathbb{F}_p$ is defined by a cubic equation $y^2 =
x^3+ax+b$ as the set

\[ \mathcal{E} = \left\{(x,y) \in \mathbb{F}_p \times \mathbb{F}_p, y^2 \equiv x^3 + a x + b ~(\textsf{mod}~ p)\right\}\]

\noindent where $a,b \in \mathbb{F}_p$ are constants such that $4a^3+27 b^2 \nequiv 0 ~(\textsf{mod} ~p)$. An elliptic curve over
$\mathbb{F}_p$ consists of the set of all pairs of affine coordinates
$(x,y)$ for $x,y \in \mathbb{F}_p$ that satisfy an equation of the
above form and an infinity point $\mathcal{O}$.

%
The point addition and its special case, point doubling over
$\mathcal{E}$ is defined as follows (the arithmetic operations are
defined in $\mathbb{F}_p$)~\cite{74} :

Let $P=(x_1,y_1)$ and $Q=(x_2,y_2)$ be two points of
$\mathcal{E}$. Then:

\begin{equation}
  P+Q  = \left\{
          \begin{array}{ll}
            \mathcal{O}& \qquad \text{if} \quad x_2=x_1 \text{ and } y_2=-y_1\text{,}\\
            (x_3,y_3) & \qquad \text{otherwise}.\\
          \end{array}
        \right.
\end{equation}

\begin{algorithm}[h]
\begin{small}
\caption{Keys generation program in Python/Sage.}
\begin{algorithmic}[1]
         
 \STATE   def GG1e(n):
 \STATE   ~~~~ l = 1
 \STATE   ~~~~ p = l*n-1
 \STATE   ~~~~ while not isprime(p) or not p\%3 == 2:
 \STATE   ~~~~ ~~~~ l += 1
 \STATE   ~~~~ ~~~~ p += n
 \STATE   ~~~~ F = GF(p)
 \STATE   ~~~~ H = EllipticCurve(F, [0, 1])
 \STATE   ~~~~ X = H.gen(0)
 \STATE   ~~~~ g = l*X
 \STATE   ~~~~ G,y = [],g
 \STATE   ~~~~ flag = True
 \STATE   ~~~~ while flag:
 \STATE   ~~~~ ~~~~ y = randint(0,n-1)*g
 \STATE   ~~~~ ~~~~ if y.order() == n:
 \STATE   ~~~~ ~~~~ ~~~~ G.append(y)
 \STATE   ~~~~ ~~~~ ~~~~ if len(G) == 2:
 \STATE   ~~~~ ~~~~ ~~~~ ~~~~ flag = False
 \STATE   ~~~~ return G,p
 \STATE
 \STATE def G(t):
 \STATE ~~~~ q1 = generatePrime(t)
 \STATE ~~~~ q2 = generatePrime(t)
 \STATE ~~~~ n = q1*q2
 \STATE ~~~~ GG,p = GG1e(n)
 \STATE ~~~~ return (q1,q2,GG,p)
 \STATE
 \STATE def KeyGen(bits):
 \STATE ~~~~ (q1,q2,GG,p) = G(bits)
 \STATE ~~~~ n = q1*q2
 \STATE ~~~~ g,u = GG
 \STATE ~~~~ h = q2*u
 \STATE ~~~~ return ((n,G,g,h,p),q1)
\end{algorithmic}
\label{Prog:keys generation}
\end{small}
\end{algorithm}

 where:
\begin{itemize}
\item $x_3 = \lambda^2-x_1-x_2$,
\item $y_3 = \lambda \times(x_1-x_3)-y_1$,
\end{itemize}
\begin{equation} \lambda  = \left\{
          \begin{array}{ll}
            (y_2-y_1) \times (x_2-x_1)^{-1}& \qquad \text{if} \quad P \neq Q\ ,\\
            (3 x_1^2+a) \times (2 y_1)^{-1} & \qquad \text{if} \quad P = Q.\\
          \end{array}
        \right.
\end{equation}

Finally, we define $P+\mathcal{O} = \mathcal{O} + P = P, \forall P \in
\mathcal{E}$, which leads to an abelian group $(\mathcal{E},+)$. On
the other hand the multiplication $n \times P$ means $P+P+....+P$ $n$
times and $-P$ is the symmetric of $P$ for the group law + defined
above for all $P \in \mathcal{E}$.

\subsection{Public/Private Keys Generation with ECC}
\label{subsec:generate the keys}

In this section we show how we can generate the public and private
keys for encryption, following the cryptosystem proposed by Boneh
\emph{et al.}~\cite{80}. The analysis of the complexity will be
treated in a later section.

Let $\tau > 0$ be an integer called ``security parameter''. To
generate public and private keys, first of all, two $\tau$-bits prime
numbers must be computed. Therefore, a cryptographic pseudo-random
generator can be used to obtain two vectors of $\tau$ bits, $q_1$ and
$q_2$. Then, a Miller-Rabin test can be applied for testing the
primality or not of $q_1$ and $q_2$. We denote by $n$ the product of
$q_1$ and $q_2$, $n = q_1 q_2$, and by $l$ the smallest positive
integer such that $p=l \times n - 1$. $l$ is a prime number while $p =
2 ~(\textsf{mod} ~3)$.

In order to find the private and public keys, we define a group $H$,
which presents the points of the super-singular elliptic curve $y^2 =
x^3+1$ defined over $\mathbb{F}_p$. It consists of $p+1 = n \times l$
points, and thus has a subgroup of order $n$, we call it $G$. In
another step, we compute $g$ and $u$ as two generators of $G$ and $h =
q_2 \times u$. Then, following~\cite{80}, the public key will be
presented by $(n,G,g,h)$ and the private key by $q_1$.

To give illustration of such a keys generation, a program is presented in Algorithm \ref{Prog:keys generation}.
It is written with the Python 2.6 language and the Sage library to manipulate elliptic curves.
The \textit{randint(a,b)} function is provided by the random library; it generates an integer randomly picked into the interval $[a,b]$.
The \textit{generatePrime(n)} function is not detailed here.
It receives an integer $n$ as its input argument and generates a prime number of $n$ bits.

\subsection{Encryption and Decryption}
After the private/public keys generation, we proceed now to the two
encryption and decryption phases:

\begin{itemize}
\item {\bf Encryption :} Assuming that our messages space consists of
  integers in the set $\{0,1,...,T\}$, where $T < q_2$, and $m$ the
  (integer) message to encrypt. Firstly, a random positive integer is
  picked from the interval $[0,n-1]$. Then, the cypher-text is defined
  by
  \[C = m \times g +r \times h \in G,\] in which $+$ and $\times$ refer to the
  addition and multiplication laws defined previously.

\item {\bf Decryption :} Once the message $C$ arrived to destination,
  to decrypt it, we use the private key $q_1$ and the discrete logarithm of
  $(q_1 \times C)$ base $q_1 \times g$ as follows:

\[m = \log_{q_1 \times g}{q_1 \times C.}\]

This takes expected time $\sqrt{T}$ using Pollard's lambda
method. Moreover, this decryption can be speed-up by precomputing a
table of powers of $q_1 \times g$.
\end{itemize}

\begin{algorithm}[h]
\begin{small}
\caption{Python program for Encryption and Decryption}
\begin{algorithmic}[1]
\STATE def Encrypt(Kp, M):
\STATE ~~~~ (n,G,g,h,p) = Kp
\STATE ~~~~ r = randint(0,n-1)
\STATE ~~~~ return M*g+r*h
\STATE
\STATE def Decrypt(Kp,Ks,C):
\STATE ~~~~ (n,G,g,h,p), q1 = Kp, Ks
\STATE ~~~~ P = q1*g
\STATE ~~~~ return P.discrete-log(q1*C)
\STATE
\STATE def Decrypt-product(Kp,Ks,C):
\STATE ~~~~ (n,G,g,h,p), q1 = Kp, Ks
\STATE ~~~~ g1 = modified-weil(g,g,p)
\STATE ~~~~ return log(C,g1)
\end{algorithmic}
\label{algo:encrypt et decrypt}
\end{small}
\end{algorithm}

In Algorithm \ref{algo:encrypt et decrypt} is detailed an example of encryption and decryption programs in Python/Sage.
The \textit{modified\_weil} and \textit{discrete\_log} functions are provided by Sage.

\subsection{Homomorphic Properties}

As we mentioned before, our approach ensures easy
encryption/decryption without any need of extra resources. This will
be proved in the next section. Moreover, our approach supports
homomorphic properties, which gives us the ability to execute
operations on values even though they have been encrypted. Indeed, it
allows $N$ additions and one multiplication directly on cryptograms,
which prevents the decryption phase at the aggregators level and saves
nodes energy, which is crucial in sensor networks.

Additions over cypher-texts are done as follows: let $m_1$ and $m_2$
be two messages and $C_1, C_2$ their cypher-texts respectively. Then
the sum of $C_1$ and $C_2$, let's call it $C$, is represented by $$C = C_1+C_2+ r \times h,$$ where $r$ is an integer randomly chosen in
$[0,n-1]$ and $h = q_2\times u$ as presented in the previous
section. This sum operation guarantees that the decryption value of
$C$ is the sum $m_1+m_2$. The addition operation can be done several
times, which means we can do sums of encrypted sums.

The multiplication of two encrypted values and its decryption are done
as follows: let $e$ be the modified Weil pairing on the curve and $g$,
$h$ the points of $G$ as defined previously. Let us recall that this
modified Weil pairing $e$ is obtained from the Weil pairing
$E$~\cite{80},~\cite{85} by the formula: $e(P,Q) = E(x \times P, Q)$, where $x$
is a root of $X^3-1$ on $\mathbb{F}_{p^2}$. Then, the result of the
multiplication of two encrypted messages $C_1, C_2$ is given by
$$C_m=e(C_1,C_2) + r \times h_1,$$ where $h_1=e(g,h)$ and $r$ is a
random integer pick in $[1,n]$.

The decryption of $C_m$ is equal to the discrete logarithm of $q_1
\times C_m$ to the base $q_1 \times g_1$:
\[m_1 m_2 = \log_{q_1 * g_1}{(q_1 \times C_m.)}\] where $g_1 = e(g,g)$.

The decryption program of a product is given in Algorithm \ref{algo:encrypt et decrypt} whereas the addition and multiplication over cryptograms programs are given in Algorithm \ref{prog:addition and multiplication}.

\begin{algorithm}[h]
\begin{small}
\caption{Python/Sage program of homomorphic operations}
\begin{algorithmic}[1]
 \STATE def multiply(Kp,cr1,cr2):
 \STATE ~~~~ (n,G,g,h,p) = Kp
 \STATE ~~~~ r = randint(0,n-1)
 \STATE ~~~~return modified-weil(cr1,cr2,p)\
 \STATE ~~~~~~~~~~ +r*modified-weil(g,h,p)
 \STATE
 \STATE def add(Kp,cr1,cr2):
 \STATE ~~~~ (n,G,g,h,p) = Kp
 \STATE ~~~~ r = randint(0,n-1)
 \STATE ~~~~ return cr1+cr2+r*h
\end{algorithmic}
\label{prog:addition and multiplication}
\end{small}
\end{algorithm}

\subsection{Encryption for Sensor Networks}
\label{sec:SN}

\paragraph{Our contribution compared to existing secure data aggregation}
In previous secure data aggregation protocols,
security and data aggregation are almost always achieved together in a
hop-by-hop manner. That is, data aggregators must decrypt
every message they receive, aggregate the messages
according to the corresponding aggregation function, and
encrypt the aggregation result before forwarding it. Therefore, these techniques cannot provide
data confidentiality at data aggregators and result in
latency because of the decryption/encryption process.

\begin{figure}[h]
\centering
        \includegraphics[height=5 cm]{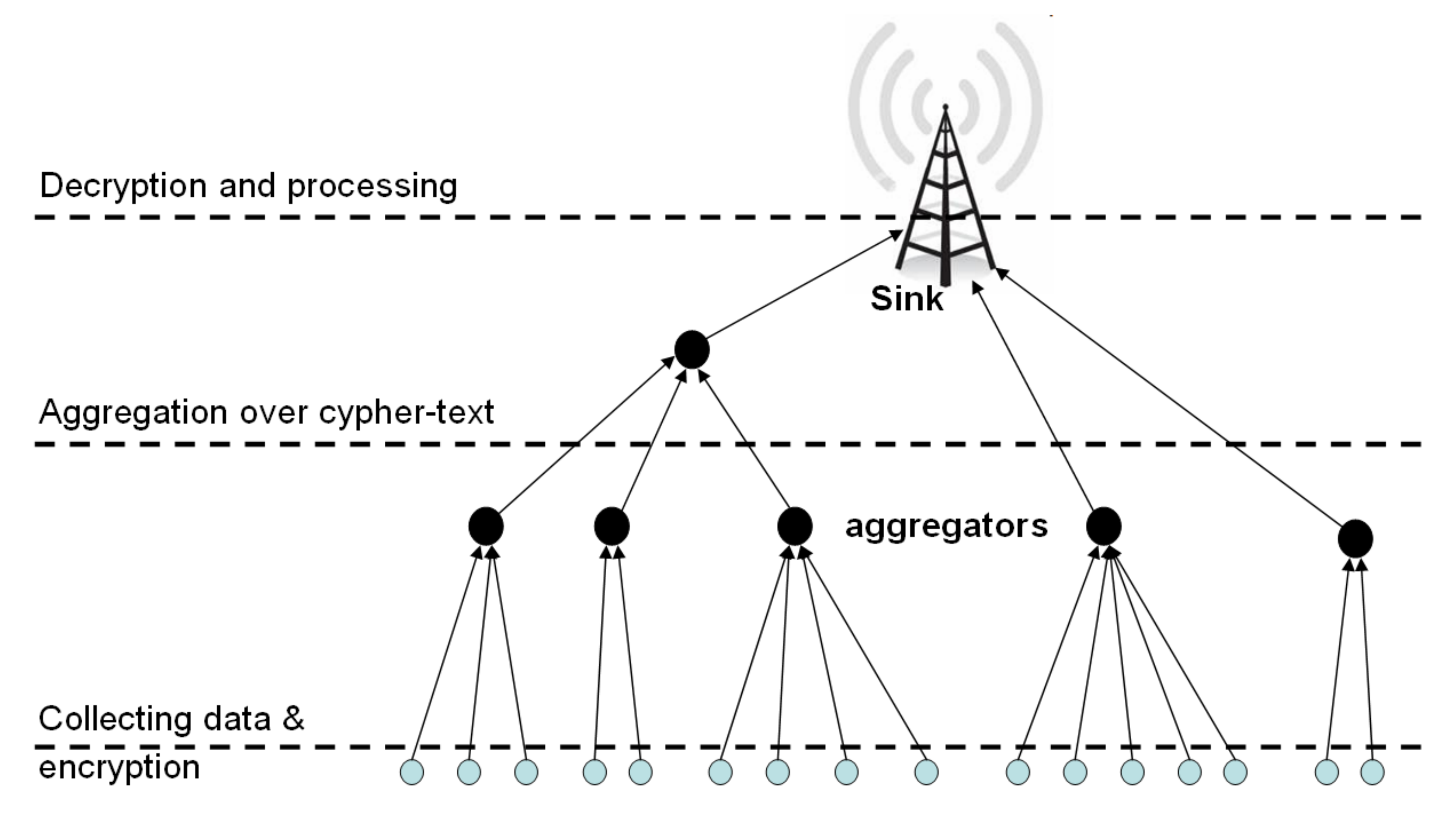}
\caption{Secure data aggregation in sensor networks}
\label{fig:agreg}
\end{figure}

In our work, we propose an encryption protocol that performs data
aggregation without requiring the decryption of the sensor
data at data aggregators. We adopt the following scenario as shown in
Figure~\ref{fig:agreg}: after collecting information, each sensor node
encrypts its data according to elliptic curve encryption (\emph{c.f.}
Section~\ref{sec:ECC}) and sends it to the nearest aggregator. Then,
aggregators aggregate the received encrypted data (without decryption)
and send it to the base station, which in his turn decrypts the data
and aggregates it. We notice that all aggregators can do $N$
additions and the final layer of aggregators can do one
multiplication on encrypted data.

\subsubsection{Illustrative Examples}

\begin{itemize}

\item {\bf Computing the Arithmetic Mean}

The arithmetic mean is the "standard" average, often simply called the "mean", defined for $n$ values $x_1, \hdots, x_n$ by

$$\bar{x} = \frac{1}{n}\cdot \sum_{i=1}^n{x_i}.$$

To compute the average of nodes measurements, aggregators can calculate the sum of the encrypted measurements and the number of nodes took these measurements and send it to the base station. More precisely, when using our scheme, each sensor encrypts its data $x_i$ to obtain $cx_i$. The sensor then forwards $cx_i$ to its parent, who aggregates all the $cx_j$'s of its $k$ children by simply adding them up. The resulting value and the encryption of $k$ are then forwarded. The sink can thus compute the average value with all of these data.

\item {\bf Computing the Variance}

Another common aggregation is to estimate the variance of the sensed values. Let us recall that the variance of $n$ values $x_1, ..., x_n$ is defined by:

$$ s_n^2 = \frac 1n \sum_{i=1}^n \left(x_i - \overline{x} \right)^ 2 = \left(\frac{1}{n} \sum_{i=1}^{n}x_i^2\right) - \overline{x}^2. $$

Our scheme can also be used to derive the variance of the measured and encrypted data, by the same method as in ~\cite{1525858}. In this case, each sensor $i$ must compute $y_i = x_i^2$, where $x_i$ is the measured sample, and encrypts $y_i$ to obtain $cy_i$. $x_i$ must also be encrypted, as explained in the previous section. The sensor forwards $cy_i$, together with $cx_i$, to its parent. The parent aggregates all the $cy_j$ of its $k$ children by simply adding them up. It also aggregates, separately, the $cx_j$, as explained in the previous section. The two resulting values are then forwarded. The sink ends up with values $Cx = \sum_{i=1}^n cx_i$ and $Cy = \sum_{i=1}^n cy_i$. $Cx$ is used to compute the average $Av$, when $Cy$ is used to compute the variance as follows: $Var = \frac{Vy}{n}-Av^2$, where $Vy$ is the decryption of $Cy$.

\item {\bf Computing the Weighted Mean}

The weighted mean of a non-empty set of data $x_1, x_2, \dots , x_n$ with non-negative weights $w_1, w_2, \dots, w_n$, is the quantity

$$\bar{x} = \frac{w_1 x_1 + w_2 x_2 + \cdots + w_n x_n}{w_1 + w_2 + \cdots + w_n}. $$

We suppose now that each aggregator $i$ of the first aggregation layer has computed the mean $x_i$ of the encrypted values received from its sensor node. Additionally, we suppose that these aggregators are weighted, depending on their importance. For security reasons, this weight is also encrypted and the cypher value is denoted by $w_i$. This $w_i$ can be proportional to the number of aggregated sensors. This weight can also illustrate the fact that two given regions have not the same relevance. To achieve weighted mean, each aggregator multiplies its encrypted mean $x_i$ with encrypted weight $w_i$ as it has been explained previously. The resulting value is then forwarded to the sink, which can decrypt $w_i \times x_i$ and sum all these decrypted values, to obtain the weighted mean defined above.
\end{itemize}

\subsection{Evaluation of the homomorphic approach}
\label{section:evaluation}

\subsubsection{Practical Issues}

In this section we present some practical issues to our data encryption
model. Firstly, we study the sizes of the encryption keys and we compare
it to existing approaches. Then, we show how we can optimize the
sizes of cryptograms in order to save more sensors energy.

\paragraph {\bf Sizes of the Keys}

Cryptograms are points of the elliptic curve $\mathcal{E}$. They are
constituted by couples of integer coordinates lesser than or equal to
$p=l q_1 q_2 - 1$.

It is commonly accepted~\cite{84},~\cite{Lenlen} that for being
secure until 2020, a cryptosystem:
\begin{itemize}
\item must have $p \approx 2^{161}$, for EC systems over $\mathbb{F}_p$,
\item must satisfy $p \approx 2^{1881}$ for classical asymmetric
  systems, such as RSA or ElGamal on $\mathbb{F}_p$.
\end{itemize}

Thus, for the same level of security, using elliptic curve cryptography
does not demand high keys sizes, contrary to the case of RSA or ElGamal
on $\mathbb{F}_p$. The use of small keys leads to small cryptograms and fast
operations for encryption.

\paragraph {\bf Reducing the Size of Cryptograms}

In this section we show how we can reduce the size of cryptograms
while using ECC. This is benefit for sensor nodes in terms of reducing
energy consumption by sending data with smaller size. The messages are
encrypted with $q_2$ bits, which leads to cryptograms with a mean of
160 bits long.

Let us suppose that $p$ $\equiv$ $3$ $($ $\textrm{mod}$ $4)$. As the
cryptogram is an element $(x,y)$ of $\mathcal{E}$, which is defined by
$y^2 = x^3+1$, we can compress this cryptogram $(x,y)$ to $(x,$ $y$
$\textrm{mod}$ $2))$ before sending it to the aggregator (as the value
of $y^2$ is known). In this situation, we obtain cryptograms with a
mean of $81$ bits long for messages between $20$ and $40$ bits long.

To decompress the cryptogram $(x,i)$, the aggregator must compute $z =
x^3 + 1 \textrm{ mod } p$ and $y=\sqrt{z}$ $\textrm{mod}$ $p$, which
can be written as $y=z^{(p+1)/4} \textrm{mod } p$, then :

\begin{itemize}
\item if $y \equiv i (\textrm{mod } 2)$, then the decompression of
  $(x,i)$ is $(x,y)$.
\item else the decompression point is $(x,p-y)$.
\end{itemize}

\subsubsection{Security study}
\label{sec:SS}

Due to hostile environments and unique characteristics of sensor
networks, it is a challenging task to protect sensitive information
transmitted by nodes to the end user. In addition, this type of
networks has security problems that traditional networks do not
face. In this section, we outline a security study dedicated to
wireless sensor networks.

In a sensor network environment adversaries can commonly use the
following attacks:

\begin{description}
\item {\bf Known-plain text attack:} They can use common key encryption
  to see when two readings are identical. By using nearby sensors
  under control, attackers can conduct a known-plain text attack.
\item {\bf Chosen-plain text attack:} Attackers can tamper with sensors
  to force them to predeterminated values.
\item {\bf Man-in-the-middle:} They can inject false readings or
  resend logged readings from legitimate sensor motes to manipulate
  the data aggregation process.
\end{description}

In Tables~\ref{Table:Polices},~\ref{Table:Polices2} and similar to~\cite{79}, we present a
comparison between different encryption policies and possible
attacks. In our method, as data are encrypted by public keys, and
these public keys are sent by the sink to the sole authenticated
motes, the wireless sensor network is then not vulnerable to a
Man-in-the-middle attacks. On the other hand, our approach guarantees
that for two similar texts gives two different cryptograms, which
prevents the Chosen-plain text attacks and the Man-in-the-middle
attacks. Finally, as the proposed scheme possesses the homomorphic
property, data aggregation is done without decryption, and no private
key is used in the network.

\begin{table*}[ht]
\centering
\caption{Encryption polices and vulnerabilities}
\renewcommand{\arraystretch}{1.3}
\label{Table:Polices}
\centering
  \begin{tabular*}{0.80\textwidth}{ll}
  \hline
  {\bf Encryption Policy} & {\bf Possible attacks}  \\
  \hline
  Sensors transmit readings without encryption & Man-in-the-middle\\
  \hline
  Sensors transmit encrypted readings  & Known-plain text attack\\
with permanent keys   & Chosen-plain text attack\\
   & Man-in-the-middle \\
    \hline
   Sensors transmit encrypted readings & None of above \\
   with dynamic keys & \\
  \hline
   Our scheme & None of above \\
  \hline
\end{tabular*}
\end{table*}

\begin{table*}[ht]
\centering
\caption{Encryption polices and aggregation}
\renewcommand{\arraystretch}{1.}
\label{Table:Polices2}
\centering
  \begin{tabular*}{0.95\textwidth}{ll}
  \hline
  {\bf Encryption Policy}  & {\bf Data aggregation} \\
  \hline
  Sensors transmit readings without encryption & Generating wrong aggregated results\\
  \hline
  Sensors transmit encrypted readings   & Data aggregation is impossible,  \\
with permanent keys   & unless the aggregator has encryption keys\\
    & \\
    \hline
   Sensors transmit encrypted readings & Data aggregation cannot be achieved\\
   with dynamic keys  & unless the aggregator has encryption keys\\
  \hline
   Our scheme  & Data aggregation can be achieved\\
  \hline
\end{tabular*}
\end{table*}

\subsection{Experimental Results}
\label{sec:simul}

\subsubsection{Simulations}

To show the effectiveness of our approach we conducted a series of
simulations comparing our method to another existing one based on RSA
cryptosystem. We considered a network formed of $500$ sensor nodes,
each one is equipped by a battery of $100$ units capacity. We consider
that the energy consumption ``$E$'' of a node is proportional to the
computational time $t$, \emph{i.e.}, $E = kt$. The same coefficient of
proportionality $k$ is taken while comparing the two encryption
scenarii. Sensor nodes are then connected to 50 aggregators chosen
randomly. Each sensor node choose the nearest aggregator. The running
of each simulation is as follows: each sensor node takes a random
value, encrypts it using one of the encryption methods then sends it
to its aggregator. Aggregators compute the sum of the encrypted
received data and send it to the sink. We compared our approach to the
known RSA public-key cryptographic algorithms, and we evaluated the
energy consumption of the network while varying the sizes of the keys
and obviously the security levels. The energy consumption is the units
of the battery used to do the encryption.

Tables \ref{table:capteursCE} and \ref{table:capteursRSA} show the
energy consumption of sensor nodes to do the encryption operations
using our encryption method and the RSA one respectively. We varied
the keys sizes and obviously the security levels. A security level is just an indicative factor of security, just to say that level 4 provides higher security level. We notice that for
the same level of security in our approach we used small keys while
saving more energy. For instance, for high security levels (4 for
example) a node using our approach needs to use a key of $167$ bits
instead of $1891$ in the case of RSA and consumes 0.1 \% of the
battery power instead of 3.63 \%.

\begin{table}
\centering
\begin{tabular}{|c|c|c|}
\hline
Security level & Size $p$ of the key & $E$ (battery units)\\
\hline
1 & 46 & 0.02 \\
2 & 85 & 0.05\% \\
3 & 125 & 0.07\\
4 & 167 & 0.10\\
\hline
\end{tabular}
\caption{Security vs energy at the nodes level using our approach}
\label{table:capteursCE}
\end{table}

\begin{table}
\centering
\begin{tabular}{|c|c|c|}
\hline
Security level & Size of the key & $E$ (battery units)\\
\hline
1 & 472 & 0.08\\
2 & 945 & 0.53\\
3 & 1416 & 1.63\\
4 & 1891 & 3.63\\
\hline
\end{tabular}
\caption{Security vs energy at the nodes level using RSA}
\label{table:capteursRSA}
\end{table}

Tables \ref{table:agregateursCE} and \ref{table:agregateursRSA} give
the energy consumption $E$ at the aggregation stage. The same
hypothesis as above have been made, the sole difference is that
aggregator nodes have a battery of 1000 units of energy. It can be
seen that the energy needed by aggregators are between 50 and 500
times more important in the RSA-based scheme, for the same level of
security.

\begin{table}
\centering
\begin{tabular}{|c|c|c|}
\hline
Security level & Size $p$ of the key & $E$ (battery units)\\
\hline
1 & 46 & 0.02 \\
2 & 85 & 0.04 \\
3 & 125 & 0.07\\
4 & 167 & 0.10\\
\hline
\end{tabular}
\caption{Security vs energy at the aggregator level using our approach}
\label{table:agregateursCE}
\end{table}

\begin{table}
\centering
\begin{tabular}{|c|c|c|}
\hline
Security level & Size of the key & $E$ (battery units)\\
\hline
1 & 472 & 1.13\\
2 & 945 & 8.09\\
3 & 1416 & 24.74\\
4 & 1891 & 56.27\\
\hline
\end{tabular}
\caption{Security vs energy at the aggregator level using RSA}
\label{table:agregateursRSA}
\end{table}

\begin{figure}[h]
\centering
        \includegraphics[height=6.5 cm]{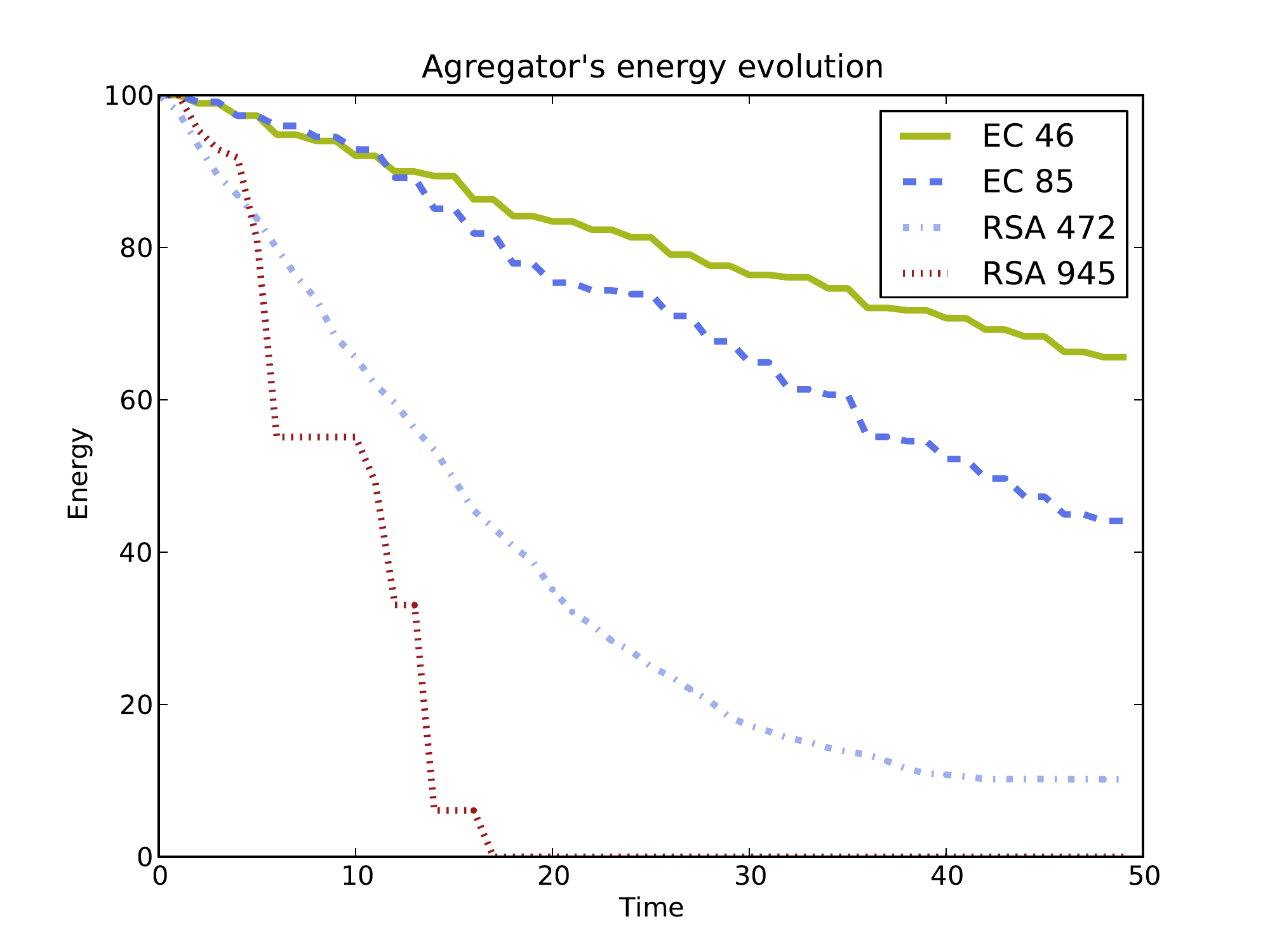}
\caption{Comparison of energy consumption}
\label{fig:reseaux}
\end{figure}

 Figure~\ref{fig:reseaux} gives the comparison between RSA and elliptic
 curve based encryption, concerning the average energy consumption of
 an aggregating wireless sensor network. We can notice that our
 approach saves the energy largely greater than the case of RSA, where
 its depletion is so fast. Finally let us notice that, in addition of
 reducing the amount of energy units needed for encryption and
 aggregation, the sink receives many more values per second in
 EC-based networks than in RSA-based one.

\section{Authentication over Homomorphic Sensor Networks}
\label{toto}

In the previous sections, we have proposed to use a homomorphism encryption scheme to support in-network processing while preserving privacy. Compared to existing secure aggregation schemes based on homomorphism encryption, our method has not been cryptanalysed. Moreover, due to the possibility to realize $n$ additions and one product over the cypher values, this scheme enlarges the variety of allowing aggregation operations through cyphertexts.

However, all of the secure homomorphism encryption schemes only allow some specific query-based aggregation functions, \emph{e.g.}, sum, average, \emph{etc.}
Indeed data encryption guarantees that only intended parties obtain the un-encrypted plain data, it does not protect the network from malicious or spoofed packets.
Node authentication enables a sensor node to ensure the identity of the packet's sender.
Another way to achieve secure data aggregation in wireless sensor networks is then to authenticate sensing values.

\begin{figure}[h]
\centering
        \includegraphics[height=5 cm]{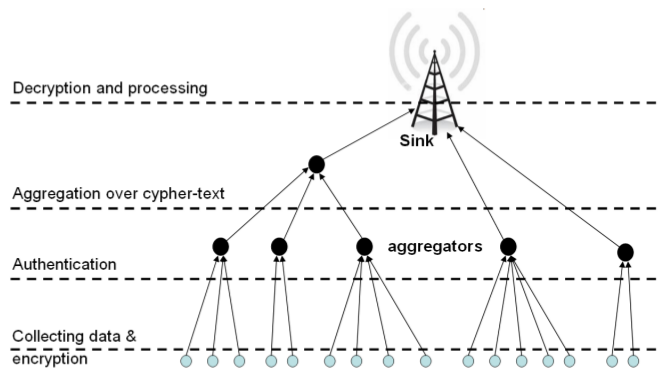}
\caption{Two layers secure data aggregation in sensor networks}
\label{fig:agreg2}
\end{figure}

Finally, an hybrid approach of secure data aggregation in wireless sensor networks can be obtained by combining homomorphic encryption and watermarking-based authentication, as it is summed up in Figure~\ref{fig:agreg2}.
In the next section we present our proposed scheme for node authentication in sensor networks.


\subsection{Information hiding-based authentication}
\label{IH based authentication}

In this paper we consider that authentication information is modulated as watermark and superposed on
the sensory data at the sensor nodes. The watermarked data can be
aggregated by the intermediate nodes without incurring any en route
checking. Upon reception of the sensory data, the data sink is
able to authenticate the data by validating the watermark, thereby
detecting whether the data has been illegitimately altered. 

Let us firstly give some recalls concerning the information hiding framework.

\subsubsection{Robustness versus Security}

Robustness and security are two major concerns in information hiding.
Even if security and robustness are neighboring concepts without clearly established definitions~\cite{Perez-Freire06}, robustness is often considered to be mostly concerned with blind elementary attacks, whereas security is not limited to certain specific attacks.
Indeed, security encompasses robustness and intentional attacks~\cite{Kalker2001,ComesanaPP05bis}.
The best attempt to give an elegant and concise definition for each of these two terms was proposed by Kalker in \cite{Kalker2001}.
Following Kalker, we will consider in this research work that:
``Robust watermarking is a mechanism to create a communication channel that is multiplexed into original content [...]. It is required that, firstly, the perceptual degradation of the marked content [...] is minimal and, secondly, that the capacity of the watermark channel degrades as a smooth function of the degradation of the marked content. [...]. Watermarking security refers to the inability by unauthorized users to have access to the raw watermarking channel. [...] to remove, detect and estimate, write or modify the raw watermarking bits.''
On the contrary, a \emph{fragile} watermarking is such that any alteration of the watermarked medium, as small as possible, will lead to the destruction of the watermark.

\subsubsection{Information hiding security}
\label{subsubsec:IH security}

In the framework of watermarking and steganography, security has seen several important developments since the last decade~\cite{BarniBF03,Cayre2005,Ker06}.
The first fundamental work in security was made by Cachin in the context of steganography~\cite{Cachin2004}.
Cachin interprets the attempts of an attacker to distinguish between an innocent image and a stego-content as a hypothesis testing problem.
In this document, the basic properties of a stegosystem are defined using the notions of entropy, mutual information, and relative entropy.
Mittelholzer, inspired by the work of Cachin, proposed the first theoretical framework for analyzing the security of a watermarking scheme~\cite{Mittelholzer99}.

These efforts to bring a theoretical framework for security in steganography and watermarking have been followed up by Kalker, who tries to clarify the concepts (robustness \emph{vs.} security), and the classifications of watermarking attacks~\cite{Kalker2001}.
This work has been deepened by Furon \emph{et al.}, who have translated Kerckhoffs' principle (Alice and Bob shall only rely on some previously shared secret for privacy), from cryptography to data hiding~\cite{Furon2002}.
They used Diffie and Hellman methodology, and Shannon's cryptographic framework~\cite{Shannon49}, to classify the watermarking attacks into categories, according to the type of information Eve has access to~\cite{Cayre2005,Perez06}, namely: Watermarked Only Attack (WOA), Known Message Attack (KMA), Known Original Attack (KOA), and Constant-Message Attack (CMA).
These categories of attacks are recalled bellow.

\begin{description}
\item[Watermark-Only Attack (WOA)] occurs when an attacker has only access to several watermarked contents.
\item[Known-Message Attack (KMA)] occurs when an attacker has access to several pairs of watermarked contents and
corresponding hidden messages.
\item[Known-Original Attack (KOA)] is when an attacker has access to several pairs of watermarked contents and
their corresponding original versions.
\item[Constant-Message Attack (CMA)] occurs when the attacker observes several watermarked contents and only knows that the
unknown hidden message is the same in all contents.
\end{description}

 Levels of security have been recently defined in these setups.
The highest level of security in WOA is called stego-security \cite{Cayre2008}, whereas chaos-security tends to improve the ability to withstand attacks in KMA, KOA, and CMA setups \cite{gfb10:ip}.

\subsubsection{Stego-security and chaos-security}

\paragraph{Stego-security}

In the prisoner problem of Simmons~\cite{Simmons83}, Alice and Bob are in jail, and they want to, possibly, devise an escape plan by exchanging hidden messages in innocent-looking cover contents.
These messages are to be conveyed to one another by a common warden, Eve, who over-drops all contents and can choose to interrupt the communication if they appear to be stego-contents.

The stego-security, defined in this
framework, is the highest security level in WOA setup~\cite{Cayre2008}.
To recall it, we need the following notations:
\begin{itemize}
  \item $\mathds{K}$ is the set of embedding keys,
    \item $p(X)$ is the probabilistic model of $N_0$ initial host contents,
      \item $p(Y|K)$ is the probabilistic model of $N_0$ watermarked contents with the same embedding key $K \in \mathds{K}$.
      \end{itemize}

In this framework, it is then supposed in this context that each host content has been watermarked with the same secret key $K$ and the same embedding function $e$.

      It is now possible to define the notion of stego-security:

      \begin{definition}[Stego-Security]
      \label{Def:Stego-security}
      The embedding function $e$ is \emph{stego-secure} if and only if:
      $$\forall K \in \mathds{K}, p(Y|K)=p(X).$$
      \end{definition}

\paragraph{Chaos-security}
\label{sec:chaos-security}

We finally detail the highest level of security in KMA, KOA, and CMA setups. It is called chaos-security and is defined as follows.

To check whether an information hiding scheme $S$ is chaos-secure or not, $S$ must be written as an iterate process $x^{n+1}=f(x^n)$ on a metric space $(\mathcal{X},d)$, where the phase space $\mathcal{X}$ is the set of all
possible contents and $d$ is a distance that must be carefully chosen,
depending on the objectives to reach ($d(x,y)$ must be small iff $x$ is
undistiguishable from $y$).
 
This formulation is always possible~\cite{2010arXiv1005.0704B}.
So,

\begin{definition}[Chaos-Security]
\label{Def:chaos-security-definition}
An information hiding scheme $S$ is said to be chaos-secure on
$(\mathcal{X},d)$ if its iterative process has a chaotic
      behavior according to Devaney.
      \end{definition}

In other words, the iteration process must satisfy the following chaos properties:
\begin{itemize}
\item iterations of $f$  
are regular (\textit{i.e.}, periodic points of $f$ are dense in 
$\mathcal{X}$),
\item $f$ is topologically transitive
(\textit{i.e.}, for any pair of open sets $U,V\subset \mathcal{X}$,
there exists some natural number $k>0$  s. t. 
$f^{k}(U)\cap V\neq \varnothing $),
\item $f$ has sensitive dependence on initial conditions
(\textit{i.e.}, there exists $\delta >0$ s.t. for any $X\in \mathcal{X}$
and any neighborhood $V$ of $X$, there exist $Y\in V$ and $k\geqslant 0$
with $d(f^{k}(X), f^{k}(Y))>\delta $).
\end{itemize}

In the approach presented in \cite{gfb10:ip}, a data hiding scheme is secure if it is unpredictable.
      Its iterative process must satisfy the Devaney's chaos property and its level of chaos-security increases with the number of chaotic properties satisfied by it.

One of these chaos property, namely the expansivity, is an important quantitative measure of disorder. It is recalled below.

\begin{definition}
A function $f$ is said to have the property of \emph{expansivity} if
\begin{equation}
\exists \varepsilon >0,\forall x\neq y,\exists n\in \mathbb{N}%
,d(f^{n}(x),f^{n}(y))\geqslant \varepsilon .
\end{equation}
\end{definition}

Then $\varepsilon $ is the \emph{constant of expansivity} of $f$: an arbitrarily small error on any initial condition is \emph{always} amplified until $\varepsilon$.
It has been proven in \cite{gfb10:ip} that an information hiding scheme that is not expansive is unable to face an attacker in KOA and KMA setups \cite{gfb10:ip}.

This new concept of security for data hiding schemes has been proposed in~\cite{2010arXiv1005.0704B} as a complementary approach to the existing framework.
It contributes to the reinforcement of confidence put into existing secure data hiding schemes.
Additionally, the study of security in KMA, KOA, and CMA setups is realizable in this context.
Finally, this framework can replace stego-security in situations that are not encompassed by it.
In particular, this framework is more relevant to give evaluation of data hiding schemes claimed as chaotic.

\subsection{Application to nodes authentication}

We explain in this section why, in our opinion, the framework recalled above is useful for studying wireless sensor network authentication nodes.

\subsubsection{Robustness for authenticated wireless sensor networks}

We now adapt the concepts recalled above for nodes authentication based on information hiding techniques.
First of all, robustness means that the watermark still remains after geometric and frequency attacks.
The interest to have a robust watermarking for authentication in WSN is then twofold.

On the one hand, the network is not always fixed and can possibly evolve over time.
Nodes can be moved for various reasons, some of them can stop to transmit their sensed and watermarked data (for technical reasons, or when they have consumed all of their energies), noise can appear during transmission, and so on.
Nevertheless, the authentication capability of the whole network must be preserved into the sink, and thus the watermarking scheme used for authentication must be compliant with such alterations.
That is to say, if the wireless sensor network can be slightly altered for reasonable and natural reasons, then the authentication scheme must be robust.
Table \ref{table:Robust and WSN} gives some relationships between geometric and frequency attacks in the information hiding framework, and natural alteration of a wireless sensor network.

\begin{table}
\centering
\begin{tabular}{|c|c|}
\hline
Digital Watermarking & WSN\\
\hline
pixel & node \\
picture & network \\
\hline
zeroing attack & death of nodes\\
rotation and resize attacks & nodes displacement \\
uniform or gaussian noise & transmission errors\\
contrast attacks & unbalanced signals \\
blur attacks & signal attenuation\\
\hline
\end{tabular}
\vspace{0.5cm}
\caption{Relationship between digital watermarking and WSN}
\label{table:Robust and WSN}
\end{table}

On the other hand, some information hiding schemes are robust against image compression attacks like JPEG or JPEG2000 compressions.
Such a resistance is obtained for instance by inserting the watermark into the DCT or DCT coefficients of the image instead of using the gray level of each pixel.
The idea formerly proposed by \cite{Zhang2008658} is to use this resistance for aggregation.
Indeed, in situation of JPEG or JPEG2000 compression resistance, watermarked data sent by sensor nodes can be aggregated by using a DCT or DWT compression.
Due to the robustness of the well-chosen information hiding scheme against these attacks, the watermark still remains after such compressions, and the aggregation preserves authentication.

Furthermore, a fragile watermarking can be useful too in an information hiding based authentication of nodes into a WSN.
Let us consider for instance that an attacker adds one of his node into a given wireless sensor network that use a fragile watermarking scheme for authentication.
Additionally, we suppose that he can send corrupted sensed values without being detected, either to the aggregation layer or to the sink.
Under this situation, the ``watermarked image'' received by the sink will be such that at least one pixel (\emph{i.e.}, the corrupted node) has not a correct piece of watermark.
Due to the fragility of the scheme, this alteration will be magnified and the extracted watermark will be completely different from what was expected, leading to the detection of the attack.
Such a fragile authentication scheme can be useful too in situations where an attacker tries to modify an authorized node. In this case, as the node can embeds different pieces of watermark, the sink will be able to detect such anomaly. 

Finally, there exist some watermarking schemes that are fragile in almost all situations, but are robust against some well defined threats.
So fragile and robust properties can help to choose the best watermarking scheme for a given WSN authentication context.
For instance, some schemes can be found in the literature that are robust against JPEG attacks with small compression rate, but are fragile in all of the other types of attacks.
Such an algorithm is helpful when the network cannot evolve geographically, must use compression based aggregation, and is in an hostile environment.
As a conclusion, we can see that using a watermarking scheme for authentication through WSN is useful in many situations.

\subsubsection{Security for authenticated WSN}

As robustness, information hiding security can be useful when authenticating nodes into wireless sensor networks.
The four classes of attacks presented in Section \ref{subsubsec:IH security} can be translated to WSN security as follows.

\begin{description}
\item[Watermark-Only Attack (WOA)] occurs when an attacker has only access to several transmitted authenticated data. That is to say, he can only observe transmissions.
\item[Known-Message Attack (KMA)] occurs when an attacker has access to several pairs of watermarked contents and corresponding hidden messages. In other words, the adversary can observe the transmissions, and has find a way to make an altered node insert the attacker's watermark.
\item[Known-Original Attack (KOA)] is when an attacker has access to several pairs of watermarked contents and their corresponding original versions. That is to say, the attacker can determine which value a node has sensed and can see the resulted watermarked data sent by this node.
\item[Constant-Message Attack (CMA)] occurs when the attacker observes several watermarked contents and only knows that the unknown hidden message is the same in all contents. In that situation, the adversary can observe another time the transmitted data. Additionally, he knows that always the same watermark is used to authenticate these data.
\end{description}

Other categories of attacks can be found in the literature, such as the Estimated-Original Attack.
They all can be translated into the wireless sensor network security framework.

The Simmons' prisoner problem put into the WSN context can be translated too, as follows.
Eve observes the transmission between two nodes called Alice and Bob.
She tries to determine whether a given transmission is authenticated or not.
Obviously, this is for her the starting point of an attack to authentication.
For instance, if she is able to make the difference between authenticated and unauthenticated data, then she can:
\begin{itemize}
\item Replace unauthenticated data with her own values without being detected.
\item Concentrate her efforts on a subset of authenticated data.
\item Try to understand the differences between authenticated and unauthenticated data, with a view to forge her own ``authenticated'' values.
\item Try to determine, by using statistical models and tools, the embedding key (the piece of watermark used to authenticate these values).
\item \emph{etc.}
\end{itemize}

The stego-security means that such a separation between authenticated and unauthenticated data is impossible, as the use of any key does not change the probabilistic model of the transmitted data.
Obviously, there is a lack of security if an authentication scheme of data sending through a WSN, with an adversary being able to observe transmissions (Simmons' prisoner problem, WOA setup), is based on a watermarking algorithm that is not stego-secure.
Similar conclusions can be obtained with the chaos-security notion in the KOA, KMA, and CMA setups: if situations covered by these setups can possibly occur, then the watermarking scheme used for authentication must be chaos-secure.

To the best of our knowledge, until now, only two data hiding schemes have been used to authenticate data sending through a WSN.
The first one is a spread-spectrum technique, used in \cite{Zhang2008658}.
The second one uses chaotic iterations.

In what follows, these information hiding techniques are recalled and their security is evaluated.

\subsection{Security study of Zhang \emph{et al.} authentication scheme}
\label{section:Zhang}

As Zhang \emph{et al.} nodes authentication scheme for WSN is based on the spread-spectrum data hiding, we must firstly recall this technique before studying its security.

\subsubsection{Spread-spectrum data hiding techniques}

Let $x \in \mathds{R}^{N_v}$ be a host vector in which we want to hide a message $m\in \{0,1\}^{N_c}$. $N_c$ is the size of the hidden payload (in bits) and $N_v$ the size of the stego or host vector (in samples). A key $\mathcal{K}$ is used to initialize a PRNG (Pseudo-Random Number Generator) to obtain $N_c$ secret carries $\{u^i\}$ taken in $\mathds{R}^{N_v}$, which can be supposed to be orthonormalized. Thus in classical SS the watermark signal $w$ is constructed as follows:
$$
\displaystyle{w = \sum_{i=0}^{N_c-1} \gamma (-1)^{m^i} u^i},
$$
\noindent where $\gamma$ is a given distortion level. The watermarked signal $y$ is then defined by:
$$
y = x + w.
$$

Let us now suppose that the components of the watermark are bounded by a finite value $\mathsf{N}_b$:\linebreak max$\left(\{ w_i, i \in \llbracket 1, N_v \rrbracket \}\right) \leqslant \mathsf{N}_b$. This bound can be as large as needed, however a very large $\mathsf{N}_b$ seems to be contradictory with the aims of a data hiding scheme.
Let us consider $\mathcal{X}=\left(\left[-\mathsf{N}_b,\mathsf{N}_b\right]^{N_v}\right)^\mathds{N}\times \mathds{R}^{N_v}$ and
$$
G((S,E)) = (\sigma(S) ; i(S) + E),
$$

\noindent where $\sigma$ is the \emph{shift} function defined by $\sigma :(S^{n})_{n\in \mathds{N}}\in \left(\left[-\mathsf{N}_b,\mathsf{N}_b\right]^{N_v}\right)^\mathds{N} \rightarrow (S^{n+1})_{n\in \mathds{N}}\in  \left(\left[-\mathsf{N}_b,\mathsf{N}_b\right]^{N_v}\right)^\mathds{N} $ and the \emph{initial function} $i$ is the map which associates to a sequence, its first term:\linebreak $i:(S^{n})_{n\in \mathds{N}}\in  \left(\left[-\mathsf{N}_b,\mathsf{N}_b\right]^{N_v}\right)^\mathds{N}\rightarrow S^{0}\in  [-\mathsf{N}_b;\mathsf{N}_b]^{N_v}$. $E$ will be the vector describing the part of the host that can be altered without sensitive damages, when $S$ will give the location of the alteration at each iteration ($S$ will depend on the hidden message and the secret key).

Spread-spectrum data hiding techniques are thus the result of $N_c$ iterations of the following dynamical system:
$$
\left\{
\begin{array}{l}
X^0 \in \mathcal{X},\\
X^{n+1} = G(X^n),
\end{array}
\right.
$$

\noindent and the watermarked media is the second component of $X^{N_c}$. Indeed, the second component of $X^k$ corresponds to the host image after $k$ alterations, whereas the first component explains how to alter it another time.

Classical SS, \emph{i.e.} with BPSK modulation~\cite{Cayre2008}, is defined by $X^0 = (S^0, E^0)$ where $E^0$ is the host vector $x$ and $S^0$ is the sequence
$$
\left((-1)^{m^0} \gamma ~u^0, (-1)^{m^1} \gamma ~u^1, \hdots, (-1)^{m^{N_c-1}} \gamma ~u^{N_c-1} \right),
$$

\noindent in which $\gamma$ allows to achieve a given distortion, whereas in ISS (Improved Spread Spectrum~\cite{Malvar03}), $S^0$ is defined by
$$
\left( (-1)^{m^i} \alpha  - \lambda \dfrac{<x,u^i>}{||u^i||^2} \right)_{i=0, \hdots, N_c-1},
$$

\noindent where $\alpha$ and $\lambda$ are computed to achieve an average distortion and to minimize the error probability~\cite{Cayre2008}. Lastly, in natural watermarking NW, $S^0$ is defined by
$$
\label{NM}
\left(- \left( 1 + \eta (-1)^{m^i} \dfrac{<x,u^i>}{|<x,u^i>|} \right) \dfrac{<x,u^i>}{||u^i||^2}
\right)_{i=0, \hdots, N_c-1}.
$$

This last modulation consists in a model-based projection on the different vectors $u^i$ followed by a $\eta-$scaling along the direction of $u^i$.

Natural watermarking has been proven stego-secure when $\eta=1$, whereas all of the other spread-spectrum techniques are not stego-secure (see~\cite{Cayre2008}).
Additionally, this scheme is reputed to be not robust.
Finally, these four techniques are chaos-secure so they can be considered when facing an attacker in the CMA context \cite{gfb10:ip}.
However, as these techniques are not expansive, they are unable to face an attacker in KOA and KMA setups \cite{gfb10:ip}.

\subsubsection{Cryptanalysis of the Zhang \emph{et al.} authentication scheme}

As recalled previously, spread spectrum is known to be not robust: even if their scheme survives to a certain degree of distortion, spread-spectrum cannot face to elementary blind attack.
Furthermore, spread-spectrum data hiding techniques are only stego-secure in the "Natural Watermarking" situation~\cite{Cayre2008}.
The spread-spectrum subclass used in~\cite{Zhang2008658} is related to classical SS, \emph{i.e.} with BPSK modulation~\cite{Cayre2008}.
This subclass is neither stego-secure~\cite{Cayre2008}, nor chaos-secure~\cite{2010arXiv1005.0704B}. These lack of security allow an attacker who observes the network to access to the secret embedding key in all of the following situations: WOA, KMA, KOA, and CMA setups. 

To improve the security of the network in WOA setup, the use of Natural Watermarking instead of BPSK modulation is required~\cite{Cayre2008}. 
However, Natural Watermarking is less chaos-secure than the data hiding algorithm presented in~\cite{bg10:ij}. 
This algorithm, based on chaotic iterations, is able to withstand attacks in KMA, KOA and CMA setups~\cite{2010arXiv1005.0705G}. 
Moreover, this technique is more robust than spread-spectrum, as it is stated in~\cite{guyeux10ter}. 
To sum up, the use of the scheme proposed in~\cite{bg10:ij} improves the security and robustness of the scheme presented in~\cite{Zhang2008658}.
This algorithm is recalled in the next section and evaluated in the last one.

\subsection{Information hiding based on chaotic iterations}
\label{section:dhci}
For easy understanding, our information hiding scheme based on chaotic iteration is explained by using pictures instead of networks.
As there is an equivalency between pixels and nodes, this discussion and the evaluation of the next section holds for a wireless sensor network, mutatis mutandis.

\subsubsection{Chaotic iterations}
\label{sec:chaotic iterations}

In the sequel $S^{n}$ denotes the $n^{th}$ term of a sequence $S$ and $V_{i}$ is for
the $i^{th}$ component of a vector $V$. Finally, the following notation
is used: $\llbracket0;N\rrbracket=\{0,1,\hdots,N\}$.\newline

Let us consider a \emph{system} of a finite number $\mathsf{N}$ of elements (or
\emph{cells}), so that each cell has a boolean \emph{state}. A sequence of length
$\mathsf{N}$ of boolean states of the cells corresponds to a particular
\emph{state of the system}. A sequence that elements belong into $\llbracket
1;\mathsf{N} \rrbracket $ is called a \emph{strategy}. The set of all strategies
is denoted by $\mathbb{S}.$

\begin{definition}
\label{Def:chaotic iterations}
The set $\mathds{B}$ denoting $\{0,1\}$, let
$f:\mathds{B}^{\mathsf{N}}\longrightarrow \mathds{B}^{\mathsf{N}}$ be a function
and $S\in \mathbb{S}$ be a strategy.
The so-called \emph{chaotic iterations} are
defined by $x^0\in \mathds{B}^{\mathsf{N}}$ and $\forall (n,i) \in
\mathds{N}^{\ast} \times \llbracket0;\mathsf{N-1}\rrbracket$:
\begin{equation}
x_i^n=\left\{
\begin{array}{ll}
x_i^{n-1} & \text{ if }S^n\neq i, \\
\left(f(x^{n-1})\right)_{S^n} & \text{ if }S^n=i.\end{array}\right.
\end{equation}
\end{definition}

\subsubsection{Chaotic iterations and Devaney's chaos}
\label{sec:topological}

In this section we give outline proofs of the properties on which our nodes authentication is based. The complete theoretical framework is detailed
in~\cite{bg10:ij}.

Denote by $\Delta $ the \emph{discrete boolean metric},
$\Delta(x,y)=0\Leftrightarrow x=y.$ Given a function $f$, define the
function: $F_{f}: \llbracket1;\mathsf{N}\rrbracket\times
\mathds{B}^{\mathsf{N}} \longrightarrow \mathds{B}^{\mathsf{N}}
$ such that $F_{f}(k,E)=\left( E_{j}.\Delta (k,j)+f(E)_{k}.\overline{\Delta
(k,j)}\right)_{j\in \llbracket1;\mathsf{N}\rrbracket}$.

Let us consider the phase space
$\mathcal{X}=\llbracket1;\mathsf{N}\rrbracket^{\mathds{N}}\times
\mathds{B}^{\mathsf{N}}$ and the map $G_{f}\left( S,E\right) =\left( \sigma (S),F_{f}(i(S),E)\right)
$, where $\sigma$ is defined by $\sigma :(S^{n})_{n\in \mathds{N}}\in \mathbb{S}\rightarrow (S^{n+1})_{n\in \mathds{N}}\in \mathbb{S}$,
and $i$ is the map $i:(S^{n})_{n\in \mathds{N}}\in \mathbb{S}\rightarrow S^{0}\in
\llbracket1;\mathsf{N}\rrbracket$. So the chaotic iterations can be described by the following iterations:
$$X^{0}\in \mathcal{X}\text{ and }X^{k+1}=G_{f}(X^{k}).$$

We have defined in~\cite{bg10:ij} a new distance $d$ between two points $(S,E),(\check{S},\check{E} )\in \mathcal{X}$
by
$d((S,E);(\check{S},\check{E}))=d_{e}(E,\check{E})+d_{s}(S,\check{S}),$
where:
\begin{itemize}
\item
$\displaystyle{d_{e}(E,\check{E})}=\displaystyle{\sum_{k=1}^{\mathsf{N}}\Delta
(E_{k},\check{E}_{k})} \in \llbracket 0 ; \mathsf{N} \rrbracket$
\item
$\displaystyle{d_{s}(S,\check{S})}=\displaystyle{\dfrac{9}{\mathsf{N}}\sum_{k=1}^{\infty
}\dfrac{|S^{k}-\check{S}^{k}|}{10^{k}}} \in [0 ; 1].$
\end{itemize}

It is then proven that,

\begin{proposition}
\label{Prop:continuite} $G_f$ is a continuous function on $(\mathcal{X},d)$.
\end{proposition}

In the metric space $(\mathcal{X},d)$,
the vectorial negation $f_{0} :\  \mathbb{B}^N  \longrightarrow  \mathbb{B}^N $, $(b_1,\cdots,b_\mathsf{N})  \longmapsto (\overline{b_1},\cdots,\overline{b_\mathsf{N}})$ satisfies the three conditions for Devaney's
chaos: regularity, transitivity, and sensitivity~\cite{bg10:ij}. So,
\begin{proposition}
$G_{f_0}$ is a chaotic map on $(\mathcal{X},d)$ according to Devaney.
\end{proposition}

To explain how to use chaotic iterations for information hiding and thus for nodes authentication, we must firstly define the significance of a given coefficient.

\subsubsection{Most and least significant coefficients}
\label{sec:msc-lsc}

We first notice that into each node, the alteration of the sensed value for authentication must not be important. That is to say, terms of the original content $x$ that may be replaced by terms issued
from the watermark $y$ are less important than other: they could be changed
without be perceived as such. More generally, a
\emph{signification function}
attaches a weight to each sensed value,
depending on its position $t$.

\begin{definition}
A \emph{signification function} is a real sequence
$(u^k)^{k \in \mathds{N}}$. 
\end{definition}

\begin{example}\label{Exemple LSC}
To illustrate this notion, we use a picture representation of a given WSN.
Let us consider a set of
grayscale images stored into portable graymap format (P3-PGM):
each pixel ranges between 256 gray levels, \textit{i.e.},
is memorized with eight bits.
In that context, we consider
$u^k = 8 - (k  \mod  8)$  to be the $k$-th term of a signification function
$(u^k)^{k \in \mathds{N}}$.
Intuitively, in each group of eight bits (\textit{i.e.}, for each pixel)
the first bit has an importance equal to 8, whereas the last bit has an
importance equal to 1. This is compliant with the idea that
changing the first bit affects more the image than changing the last one.
\end{example}

\begin{definition}
\label{def:msc,lsc}
Let $(u^k)^{k \in \mathds{N}}$ be a signification function,
$m$ and $M$ be two reals s.t. $m < M$.
\begin{itemize}
\item The \emph{most significant coefficients (MSCs)} of $x$ is the finite
  vector  $$u_M = \left( k ~ \big|~ k \in \mathds{N} \textrm{ and } u^k
    \geqslant M \textrm{ and }  k \le \mid x \mid \right);$$
 \item The \emph{least significant coefficients (LSCs)} of $x$ is the
finite vector
$$u_m = \left( k ~ \big|~ k \in \mathds{N} \textrm{ and } u^k
  \le m \textrm{ and }  k \le \mid x \mid \right);$$
 \item The \emph{passive coefficients} of $x$ is the finite vector
   $$u_p = \left( k ~ \big|~ k \in \mathds{N} \textrm{ and }
u^k \in ]m;M[ \textrm{ and }  k \le \mid x \mid \right).$$
 \end{itemize}
 \end{definition}

\begin{figure}[htb]

\begin{minipage}[b]{.98\linewidth}
  \centering
  \centerline{\includegraphics[width=3.cm]{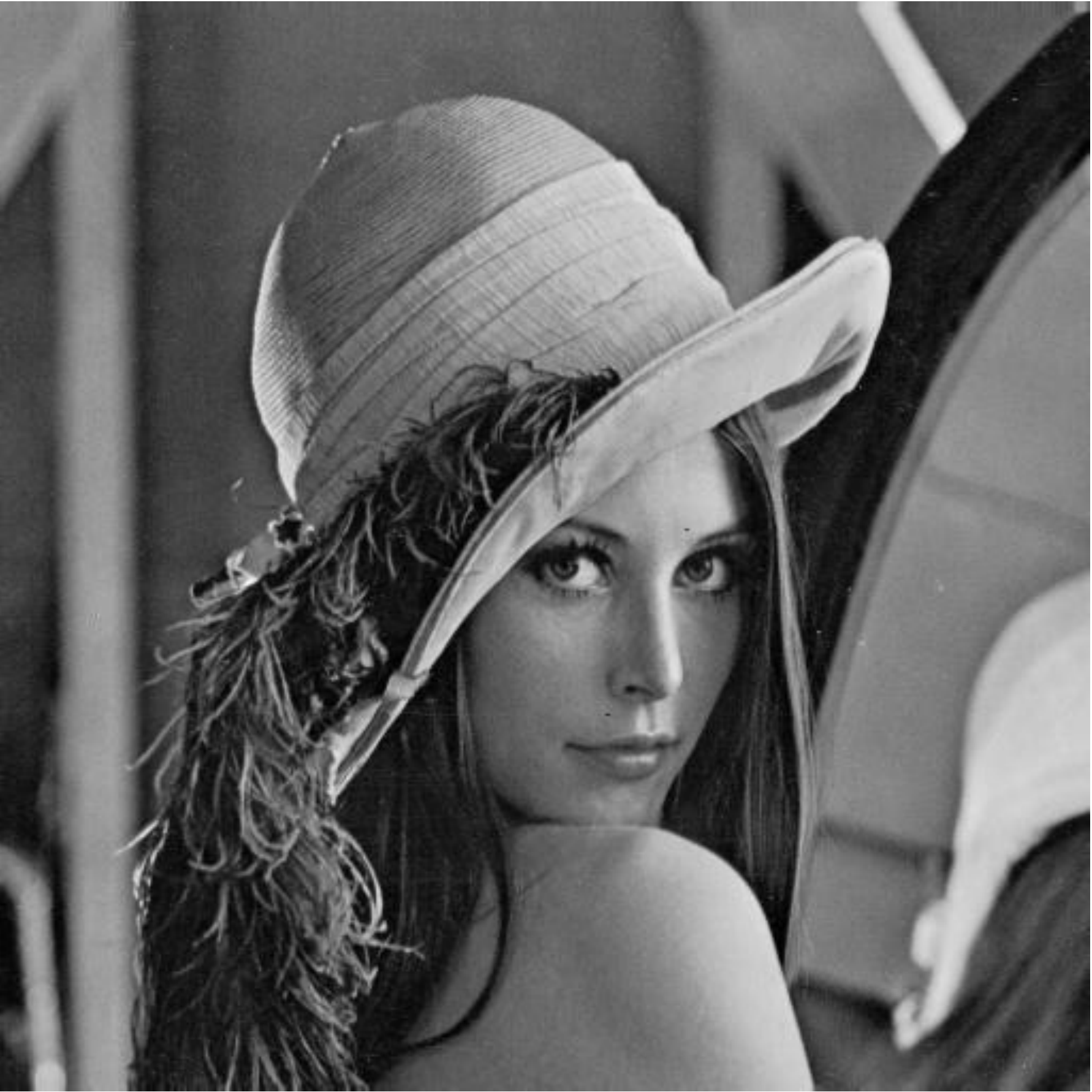}}
  \centerline{(a) Sensed values of a WSN.}
\end{minipage}
\begin{minipage}[b]{.49\linewidth}
  \centering
    \centerline{\includegraphics[width=3.cm]{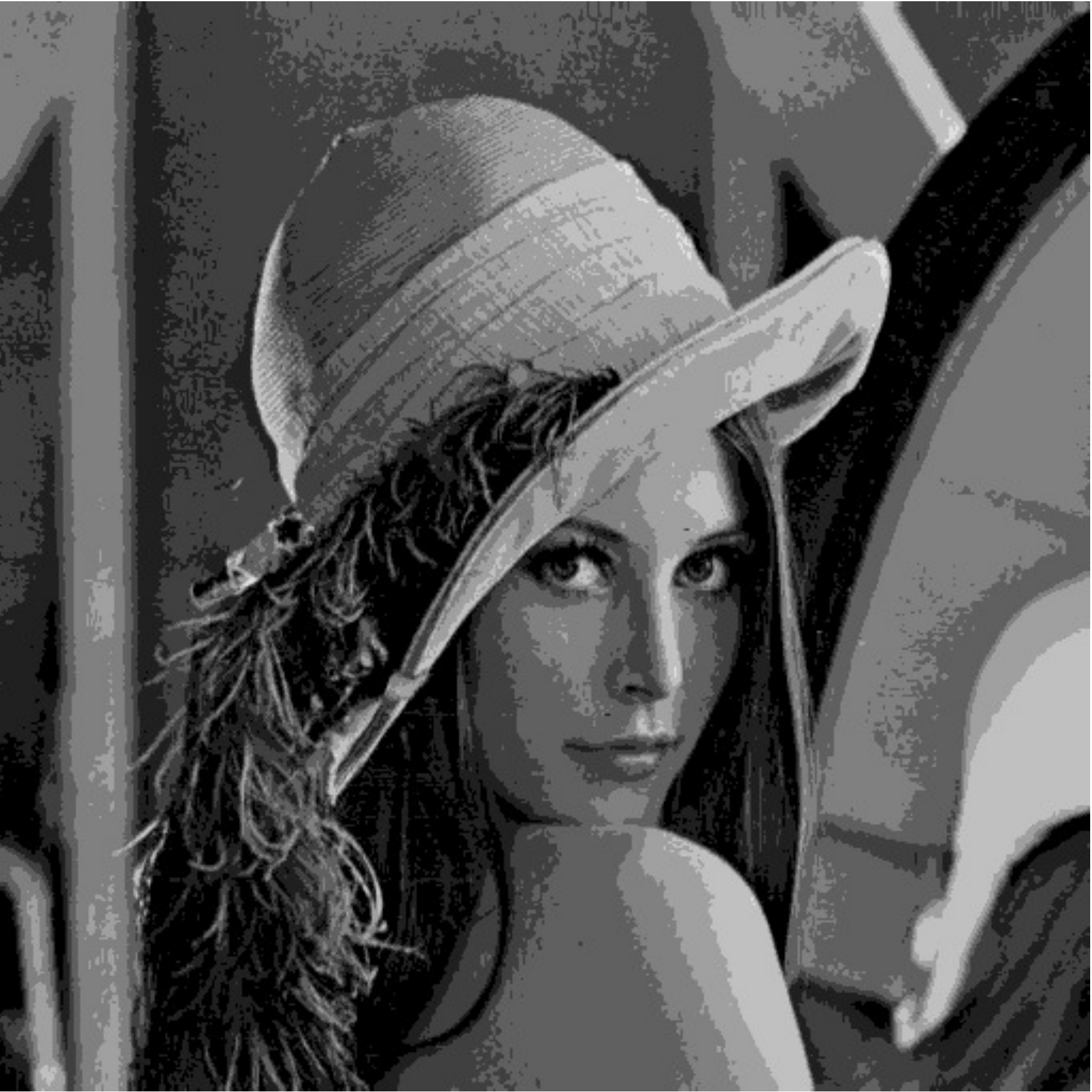}}
  \centerline{(b) MSCs of the WSN.}
\end{minipage}
\hfill
\begin{minipage}[b]{0.49\linewidth}
  \centering
    \centerline{\includegraphics[width=3.cm]{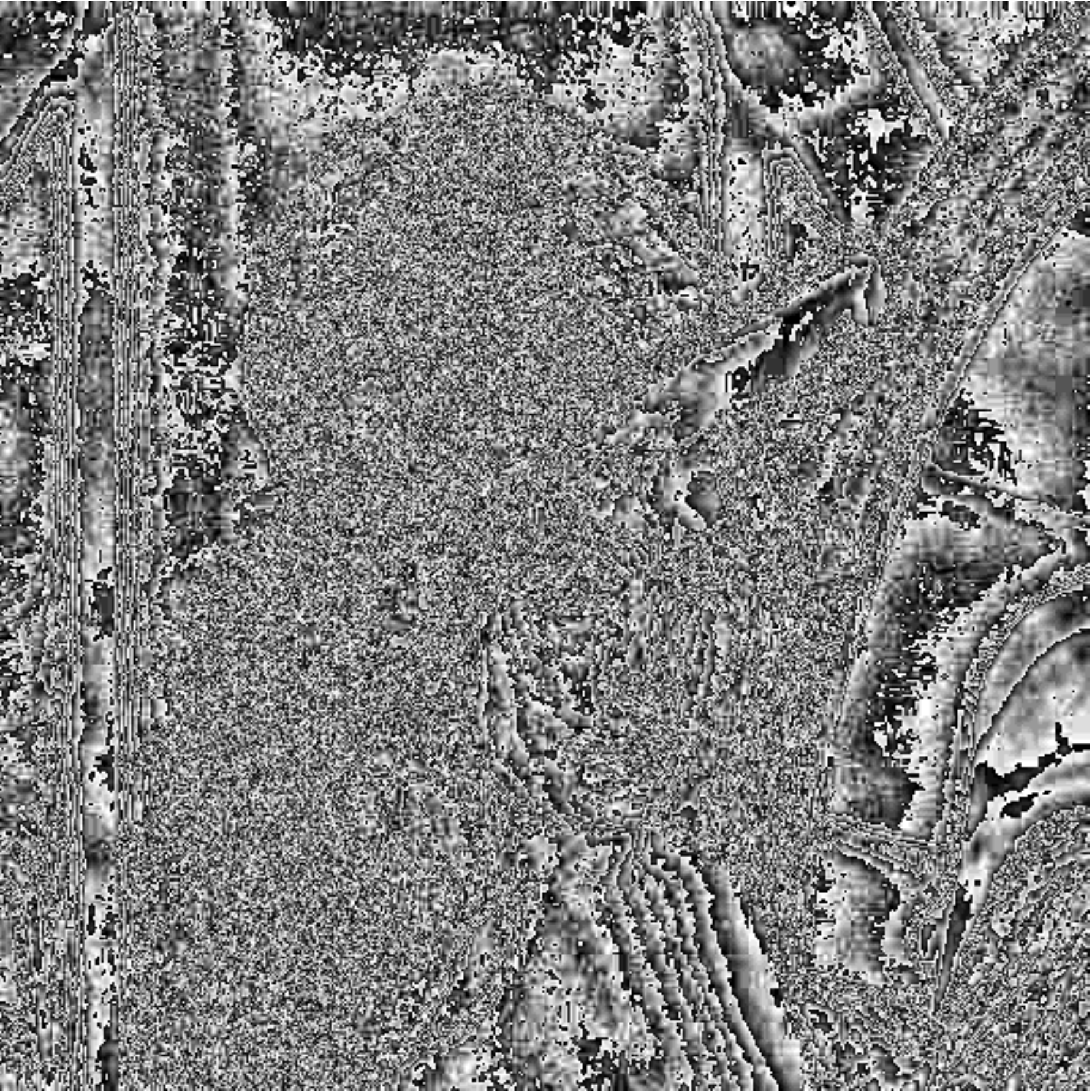}}

  \centerline{(c) LSCs of the WSN ($\times 17$).}
\end{minipage}
\caption{Most and least significant coefficients of the WSN.}
\label{fig:MSCLSC}
\end{figure}

For a given WSN $x$,
MSCs are then ranks of $x$  that describe the relevant part
of the sensed values, whereas LSCs translate its less significant parts.
These two definitions are illustrated on Figure~\ref{fig:MSCLSC}, where the significance function $(u^k)$ is defined as in Example \ref{Exemple LSC}, $M=5$, and $m=6$.

\subsubsection{Presentation of the scheme}

Authors of \cite{guyeux10ter} have proposed to use chaotic iterations as an information hiding scheme, as follows.
Let:

\begin{itemize}
  \item $(K,N) \in [0,1]\times \mathds{N}$ be an embedding key,
  \item $X \in \mathbb{B}^\mathsf{N}$ be the $\mathsf{N}$ LSCs of a cover $C$,
  \item $(S^n)_{n \in \mathds{N}} \in \llbracket 1, \mathsf{N} \rrbracket^{\mathds{N}}$ be a strategy, which depends on the message to hide $M \in [0,1]$ and $K$,
  \item $f_0 : \mathbb{B}^\mathsf{N} \rightarrow \mathbb{B}^\mathsf{N}$ be the vectorial logical negation.
\end{itemize}

So the watermarked media is $C$ whose LSCs are replaced by $Y_K=X^{N}$, where:

\begin{equation}
\left\{
 \begin{array}{l}
X^0 = X\\
\forall n < N, X^{n+1} = G_{f_0}\left(X^n\right).\\
\end{array} \right.
\end{equation}

To sum up, chaotic iterations are realized on the least significant part of the sensed values.

Two ways to generate $(S^n)_{n \in \mathds{N}}$ are given by these authors, namely
Chaotic Iterations with Independent Strategy~(CIIS) and Chaotic Iterations with Dependent
Strategy~(CIDS).
In CIIS, the strategy is independent from the cover media $C$, whereas in CIDS the strategy will be dependent on $C$.
As we will use the CIIS strategy in this document, we recall it below.
Finally, MSCs can be used as a parameter of the CIIS.

\subsubsection{CIIS strategy}

Let us firstly give the definition of the Piecewise Linear Chaotic Map~(PLCM, see~\cite{Shujun1}):

\begin{equation}
F(x,p)=\left\{
 \begin{array}{ccc}
x/p & \text{if} & x \in [0;p], \\
(x-p)/(\frac{1}{2} - p) & \text{if} & x \in \left[ p; \frac{1}{2} \right],
\\
F(1-x,p) & \text{else,} & \\
\end{array} \right.
\end{equation}

\noindent where $p \in \left] 0; \frac{1}{2} \right[$ is a ``control parameter''.

Then, the general term of the strategy $(S^n)_n$ in CIIS setup is defined by
the following expression: $S^n = \left \lfloor \mathsf{N} \times K^n \right \rfloor +
1$, where:

\begin{equation}
\left\{
 \begin{array}{l}
p \in \left[ 0 ; \frac{1}{2} \right] \\
K^0 = M \otimes K\\
K^{n+1} = F(K^n,p), \forall n \leq N_0\\ \end{array} \right.
\end{equation}

\noindent in which $\otimes$ denotes the bitwise exclusive or (XOR) between two floating part numbers (\emph{i.e.}, between their binary digits representation), $K$ is (one of) the embedding key, and $M$ is:
\begin{itemize}
\item either the sequence of MSCs, in the particular situation where each node has access to the other ones and when a fragile watermarking is needed ({\it Authentication}),
\item or the rest of the embedding key, when robustness is required ({\it Unauthentication}).
\end{itemize}

\subsection{Our proposed method}

To prove the efficiency and the robustness of the proposed algorithm, some
attacks are applied to our chaotic watermarked image. For each attack, a
similarity percentage with the watermark is computed, this percentage is the
number of equal bits between the original and the extracted watermark.
These results have been formerly obtained in \cite{guyeux10ter}.

\subsubsection{Zeroing Attack}

\noindent In this kind of attack, some nodes of the WSN are put to 0. In this case, the results in Table~\ref{Table:Crop} have been obtained. We can conclude that in case of unauthentication, the
watermark still remains after a cropping attack: the desired robustness is
reached. In case of authentication, even a small change of the carrier sensed values lead to a very different extracted watermark.
In this case, any attempt to alter the WSN will be signaled.

\begin{table}
\centering
\begin{tabular}{|c|c||c|c|}
\hline
\multicolumn{2}{|c||}{UNAUTHENTICATION} & \multicolumn{2}{c|}{AUTHENTICATION}
\\ \hline
Size (pixels) & Similarity & Size (pixels) & Similarity \\ \hline
10 & 99.08\% & 10 & 89.81\% \\
50 & 97.31\% & 50 & 54.54\% \\
100 & 92.43\% & 100 & 52.24\% \\ \hline
\end{tabular}
\caption{Zeroing attacks.}
\label{Table:Crop}
\end{table}

\subsubsection{Rotation Attack}

\noindent Let $r_{\theta }$ be the rotation of angle $\theta$ around the center $(128, 128)$ of the carrier image. So, the transformation $r_{-\theta }\circ r_{\theta }$ is applied to the watermarked WSN.
The good results in Table~\ref{tab:rot} are obtained.

\begin{table}
\centering
\begin{tabular}{|c|c||c|c|}
\hline
\multicolumn{2}{|c||}{UNAUTHENTICATION} & \multicolumn{2}{c|}{AUTHENTICATION}
\\ \hline
Angle  & Similarity & Angle & Similarity \\ \hline
5° & 94.67\% & 5° & 59.47\% \\
10° & 91.30\% & 10° & 54.51\% \\
25° & 80.85\% & 25° & 50.21\% \\ \hline
\end{tabular}
\caption{Rotation attacks.}
\label{tab:rot}
\end{table}

\subsubsection{JPEG Compression}

A JPEG compression is applied to the sensed values, depending on a
compression level. Let us notice that this attack leads to a change of
the representation domain (from spatial to DCT domain). In this case, the
results in Table~\ref{tab:jpeg} have been found.
A good authentication through a compression-based aggregation is obtained. As for the
unauthentication case, the watermark still remains after a compression level
equal to 10. This is a good result if we take into account the fact that we
use ``spatial'' embedding.

\begin{table}
\centering
\begin{tabular}{|c|c||c|c|}
\hline
\multicolumn{2}{|c||}{UNAUTHENTICATION} & \multicolumn{2}{c|}{AUTHENTICATION}
\\ \hline
Ratio & Similarity & Ratio & Similarity \\ \hline
2 & 82.95\% & 2 & 54.39\% \\
5 & 65.23\% & 5 & 53.46\% \\
10 & 60.22\% & 10 & 50.14\%\\ \hline
\end{tabular}
\caption{JPEG compression attacks.}
\label{tab:jpeg}
\end{table}

\subsubsection{Gaussian Noise}

\noindent Watermarked image can be also attacked by the addition of a Gaussian noise,
depending on a standard deviation. In this case, the results in Table~\ref{tab:gau} have
been found.

\begin{table}
\centering
\begin{tabular}{|c|c||c|c|}
\hline
\multicolumn{2}{|c||}{UNAUTHENTICATION} & \multicolumn{2}{c|}{AUTHENTICATION}
\\ \hline
Standard dev. & Similarity & Standard dev. & Similarity \\ \hline
1 & 74.26\% & 1 & 52.05\% \\
2 & 63.33\% & 2 & 50.95\% \\
3 & 57.44\% & 3 & 49.65\% \\ \hline
\end{tabular}
\caption{Gaussian noise attacks.}
\label{tab:gau}
\end{table}


\section{Conclusion}
In this paper, we presented a two layers secure data aggregation for sensor networks. The first layer is based on data encryption
with homomorphic properties that provide the possibility to operate on
cypher-text. It prevents the decryption phase at the aggregators
layers and saves nodes energy. Existing works have exponential bound
in computation and are not suitable for sensor networks, which we tried
to relax in our approach. The proposed scheme permits the generation
of shorter encryption asymmetric keys, which is so important in the
case of sensor networks. The second layer proposes a watermarking-based authentication scheme. The distinct
advantage of this layer is to achieve end-to-end authentication where the sink can directly validate the
received data from the sources. The experimental results show that our method
significantly reduces computation and communication overhead compared
to other works, and can be practically implemented in on-the-shelf
sensor platforms.

\bibliographystyle{plain}
\bibliography{mabase}

\end{document}